\documentclass[12pt]{reportj}
\usepackage{longtable}
\usepackage{hyperref} 
\makeatletter
\def\@to{to}
\makeatother
\usepackage{times}
\usepackage{graphicx}
\usepackage{tocloft}
\usepackage{fancyheadings}
\emergencystretch=\maxdimen
\renewcommand\thesection{\arabic{section}}
\renewcommand\thesubsection{\thesection.\arabic{subsection}}

\def\ssection#1{\setcounter{subsection}{0} \refstepcounter{section} \section*{\hbox to \hsize{\large\bf \arabic{section}. #1\hfill }}\label{sec} \addcontentsline{toc}{section}{\arabic{section}. #1}}
\def\ssubsection#1{\setcounter{subsubsection}{0} \refstepcounter{subsection}\subsection*{\hbox to \hsize{\normalsize\bfseries\itshape \arabic{section}.\arabic{subsection} #1\hfill}}\label{subsec} \addcontentsline{toc}{subsection}{\arabic{section}.\arabic{subsection} #1}}
\def\ssubsubsection#1{\refstepcounter{subsubsection}\subsection*{\hbox to \hsize{\normalsize\it \arabic{section}.\arabic{subsection}.\arabic{subsubsection} #1\hfill}}\label{subsubsec} \addcontentsline{toc}{subsubsection}{\arabic{section}.\arabic{subsection}.\arabic{subsubsection} #1}}

\def\ssectionstar#1{\section*{\hbox to \hsize{\large\bf #1\hfill}} \addcontentsline{toc}{section}{#1}}
\def\ssubsectionstar#1{\subsection*{\hbox to \hsize{\normalsize\bfseries\itshape #1\hfill}} \addcontentsline{toc}{subsection}{#1}}
\def\ssubsubsectionstar#1{\subsection*{\hbox to \hsize{\normalsize\it  #1\hfill}} \addcontentsline{toc}{subsection}{#1}}

\renewcommand{\cftaftertoctitle}{%
\mbox{}\hfill{\normalfont Page}}
\begin{document}

~\\

\vspace{-2.4cm}
\noindent\includegraphics*[width=0.295\linewidth]{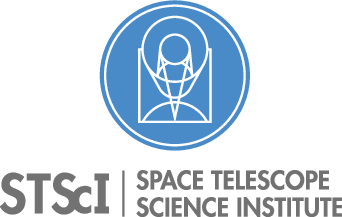}

\vspace{-0.4cm}

\begin{flushright}
    {\bf Instrument Science Report STIS 2024-01}
    
    \vspace{1.1cm}
    
    {\bf\Huge Safety Acquisitions: Redundancy for non-repeatable multi-orbit STIS visits}
    
    \rule{0.25\linewidth}{0.5pt}
    
    \vspace{0.5cm}
    
    Matthew M. Dallas$^1$, Matthew R. Siebert$^1$
    \linebreak
    \newline
    \footnotesize{$^1$ Space Telescope Science Institute, Baltimore, MD}
    
    \vspace{0.5cm}
    
     30 January 2024
\end{flushright}

\vspace{0.1cm}

\noindent\rule{\linewidth}{1.0pt}
\noindent{\bf A{\footnotesize BSTRACT}}

{\it \noindent For observations of supernovae, kilonovae, tidal disruption events, and other non-repeatable observations, it is important the science data is taken successfully within a specific time window. Part of obtaining that data is often centering objects in the aperture to a higher accuracy than is available from Hubble Space Telescope's (HST's) blind pointing. On the HST Space Telescope Imaging Spectrograph (STIS) the sequence of exposures responsible for this centering is the target acquisition or STIS ACQ sequence, and it is most often placed only at the beginning of a visit. Unfortunately, STIS ACQ sequences will fail if the observatory experiences issues locating guide stars in time for the start of the required exposures. If the guide stars are located at a later point in the visit, the remaining science exposures can be taken but the pointing might not be as accurate as is required. This work discusses both the frequency of this issue and the feasibility of placing redundant or ``safety" STIS ACQ sequences in a multi-orbit visit to regain the desired pointing accuracy in an affected visit. To do so we select a subset of all 113 STIS ACQ sequences from September 2018 to September 2023 which have experienced this issue. We find that this problem occurs in $\sim$5\% of the total STIS ACQ sequences taken during that time period, with a recent increase in the rate to $\sim$9\% from March to September 2023. Since the observatory goes through periods of better or worse pointing performance, this recent increased failure rate is not guaranteed to continue. For those failed visits which span multiple orbits, $\sim$39\% never obtain a lock on the guide stars and thus take no data. Of the multi-orbit visits that do recover the guide stars, the majority ($\sim$78\%) do so before the beginning of science exposures in the second orbit. We also provide advice for users on how to make a risk assessment based on the analysis presented here.}

\vspace{-0.1cm}
\noindent\rule{\linewidth}{1.0pt}

\renewcommand{\cftaftertoctitle}{\thispagestyle{fancy}}
\tableofcontents


\vspace{-0.3cm}
\ssection{Introduction}\label{sec:Introduction}

\begin{figure}[!h]
  \centering
  \includegraphics[width=\textwidth]{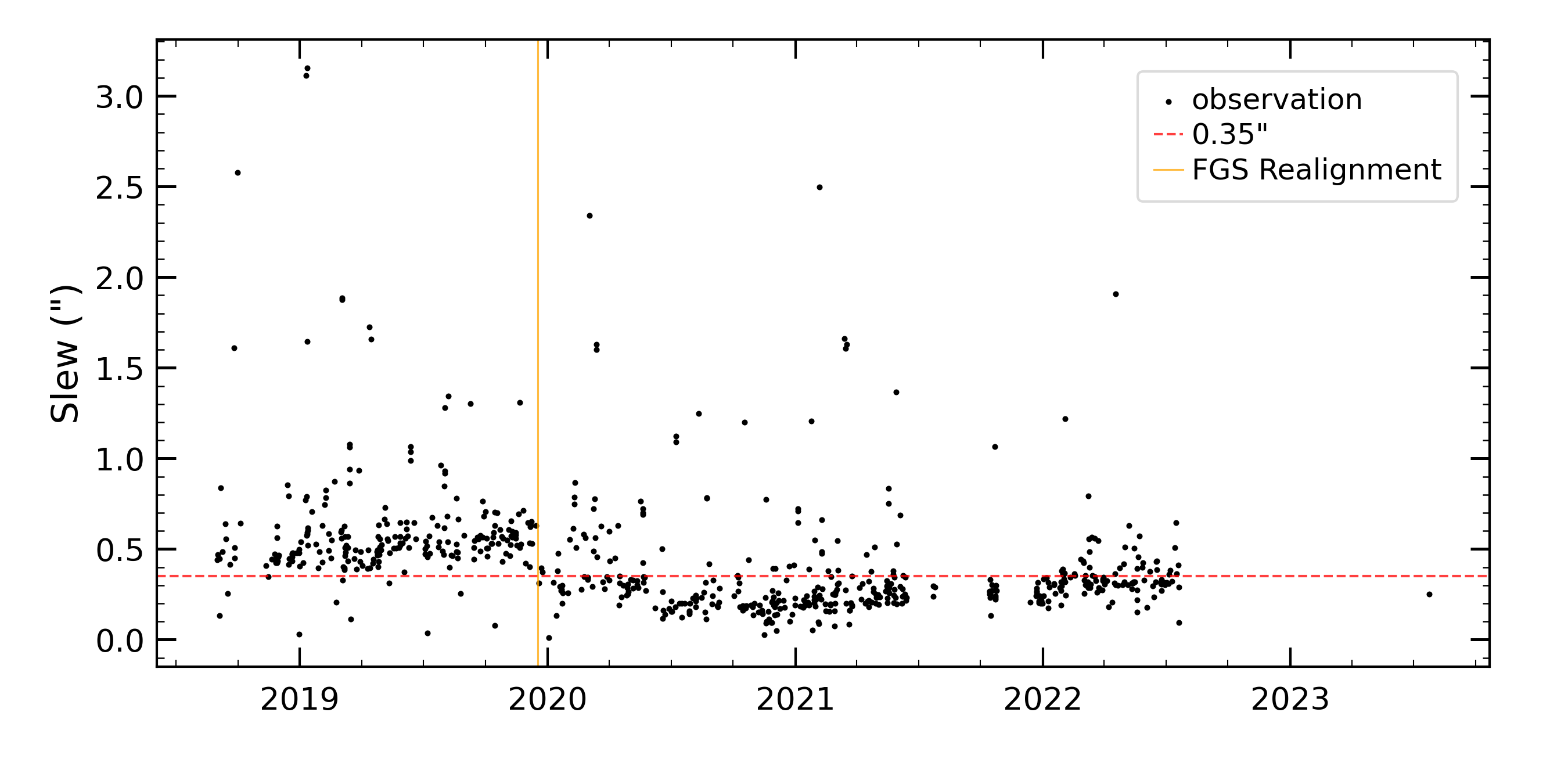}
    \caption{Amount of total slewing required to move the science target into the slit during the STIS ACQ sequence. We have selected only a subset of observations where one of the three FGSs (FGS 2) was dominant. The vertical line represents an update to the FGS-STIS alignment which improved the blind pointing accuracy. These updates occur periodically. From August 2022 onward FGS 1 and 3 have primarily been used as the dominant sensor and have blind pointing errors slightly less than 0.35 arcseconds. We show the FGS 2 dominant observations here since of the three it most clearly shows the improvement the FGS-STIS realignment made to the blind pointing accuracy.}
    \label{fig:blind_pointing}
\end{figure}

For most Space Telescope Imaging Spectrograph (STIS) observations, a high pointing accuracy is required to place objects in a scientific aperture. At the beginning of an orbit the Hubble Space Telescope (HST) acquires guide stars via the Fine Guidance Sensors (FGSs) in order to control the telescope's pointing. After acquiring guide stars the blind pointing accuracy of a science target in the STIS frame is currently $\sim$ 0.35 arcseconds. This uncertainty is due to the accuracy of the guide star catalog positions, the FGS-STIS alignment, and the target position (see STIS Instrument Handbook Section 8.1.1). To further center targets within the aperture, STIS uses a series of target acquisition images (ACQs). If guide stars are successfully acquired, the STIS ACQ sequence is performed. Upon successful completion of the STIS ACQ sequence a point source target should be centered to an accuracy $(2\sigma)$ of ~0.01 arcseconds (see STIS Instrument Handbook Section 8.1.2). The blind pointing accuracy is characterized as the total amount of slewing in arcseconds required during the STIS ACQ sequence phase. A subset of these measurements are shown as a function of time in Figure~\ref{fig:blind_pointing}. 

Initial guide star acquisitions and the subsequent STIS ACQ sequences are not always successful. During a recent unrepeatable observation of a supernova the HST guide star acquisition failed at the beginning of the first orbit, causing the shutter to remain closed for the duration of the STIS ACQ sequence and remaining science exposures of the first orbit. At the beginning of the second orbit the guide stars were located and STIS began taking data, but there was no way to ensure the target was centered in the slit since the STIS ACQ sequence had not been performed. The visit continued with four more orbits taking multiple science exposures with this pointing. The visit initiated discussions in the STIS team on how to mitigate potential science loss from missed STIS ACQ sequences such as selecting wider apertures or placing multiple STIS ACQ sequences in a visit. This latter strategy of placing additional redundant or “safety” STIS ACQ sequences after subsequent HST guide star re-acquisitions for multi-orbit, unrepeatable (e.g., SNe, KNe) or highly-constrained (e.g., exoplanet) observations may be favorable for high-priority time-sensitive targets where a substantial loss of data cannot be reconciled with a Hubble Observation Problem Report (HOPR). In such circumstances, Principle Investigators (PIs) may be inclined to trade exposure time for more insurance on the success of their observations.

The rest of this work will investigate how implementing safety ACQ sequences would impact a time sensitive observing program. These safety ACQ sequences are additional STIS ACQ sequences placed in the subsequent orbits after an HST guide star re-acquisition in order to implement the STIS ACQ sequence pointing correction in the event the first one was missed. We begin with a brief overview of the steps taken during a STIS ACQ sequence and how they can be affected by guide star issues, followed by a description of how we select a catalog of visits with failed STIS ACQs as a result of guide star issues. We then produce statistics on that catalog of guide star impacted STIS ACQs showing how often these visits occur, how often the guide stars are reacquired, and if so, where in the duration of the visit this occurs. Finally, we provide advice on how one could use this information in order to make an informed risk assessment as to whether or not their program may wish to sacrifice some exposure time to a potentially redundant additional STIS ACQ sequence.


\lhead{}
\rhead{}
\cfoot{\rm {\hspace{-1.9cm} Instrument Science Report STIS 2024-01 Page \thepage}}

\vspace{-0.3cm}
\ssection{STIS Acquisitions}\label{sec:acq}

Here we give a general overview of the parts of the STIS ACQ process. A STIS ACQ sequence comprises a series of charge-coupled device (CCD) exposures intended to accurately center an object in a given scientific aperture. The sequence includes two stages: the coarse-locate phase, and the fine-locate phase. Before the coarse-locate phase upon a successful guide star acquisition, the observatory has an initial pointing accurate to $\sim$ 0.35 arcseconds, well within STIS's widest supported aperture (2 arcsecond width), but larger than the more typically used 0.2 arcsecond width apertures. The coarse-locate phase then begins by taking two dithered 5x5 arcsecond CCD images to remove hot pixels with larger CCD images used for diffuse sources (see STIS Instrument Handbook Section 8.2.2). After the flight software does a quick reduction of the images, it calculates the pixel coordinates of the brightest source in the image. This processed image is the first science extension of an ACQ Flexible Image Transport System (FITS) file. Based on the calculated coordinates of the target on the CCD, the spacecraft moves to place the target at the nominal slit position.

The fine-locate phase then begins by re-imaging the target and re-calculating its coordinates on the CCD. This is the second science extension of an ACQ FITS file. In order to find the actual location of the slit, the external shutter is closed and the slit is illuminated by the Hole in the Mirror (HITM) lamp and an image is taken. This image is similarly processed, and the coordinates of the center of the slit are calculated. This processed HITM image is the third science extension of an ACQ FITS file. The spacecraft performs a small angle maneuver to place the target in the center of the aperture based on the difference between the calculated coordinates of the target location and the slit. After the target is fully centered a final small angle maneuver is performed to place the target in the specified scientific aperture before the science exposures begin. The STIS ACQ sequence takes roughly six minutes depending on the exposure time provided, which depends on the magnitude of the target and selected filter. For a more in-depth discussion of STIS ACQ sequences including advice on how to create an acquisition observing scheme, see STIS Instrument Handbook Section 8 and STIS Data Handbook Section 5.2. An example of the images in an ACQ FITS file for a successful STIS ACQ sequence is shown below in Figure \ref{fig:good_acq}, taken from the internal STIS ACQ sequence monitor.

\begin{figure}[!h]
  \centering
  \includegraphics[width=\textwidth]{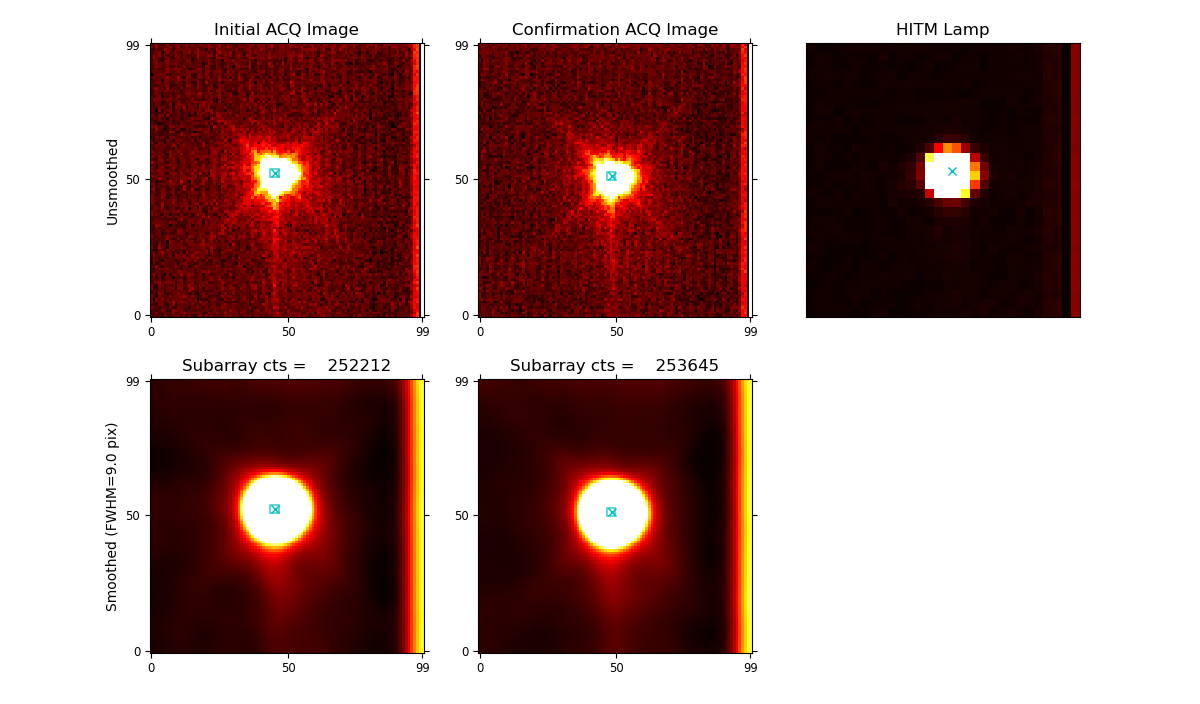}
    \caption{Images taken during a successful STIS ACQ sequence for a follow up visit in the same program discussed in Section \ref{sec:Introduction} which experienced a failed STIS ACQ sequence in a prior visit. From left to right is the coarse-locate image representing the initial pointing, followed by the fine-locate phase image taken after a slew based on the position of the target calculated in the coarse-locate phase, and finally the HITM exposure taken in order to calculate the true position of the slit. The bottom two images are Gaussian smoothed versions of the first two, titled with the sum of the counts in each image. The X overlaid in the images is at the 2D centroid.}
    \label{fig:good_acq}
\end{figure}

\begin{figure}[!h]
  \centering
  \includegraphics[width=\textwidth]{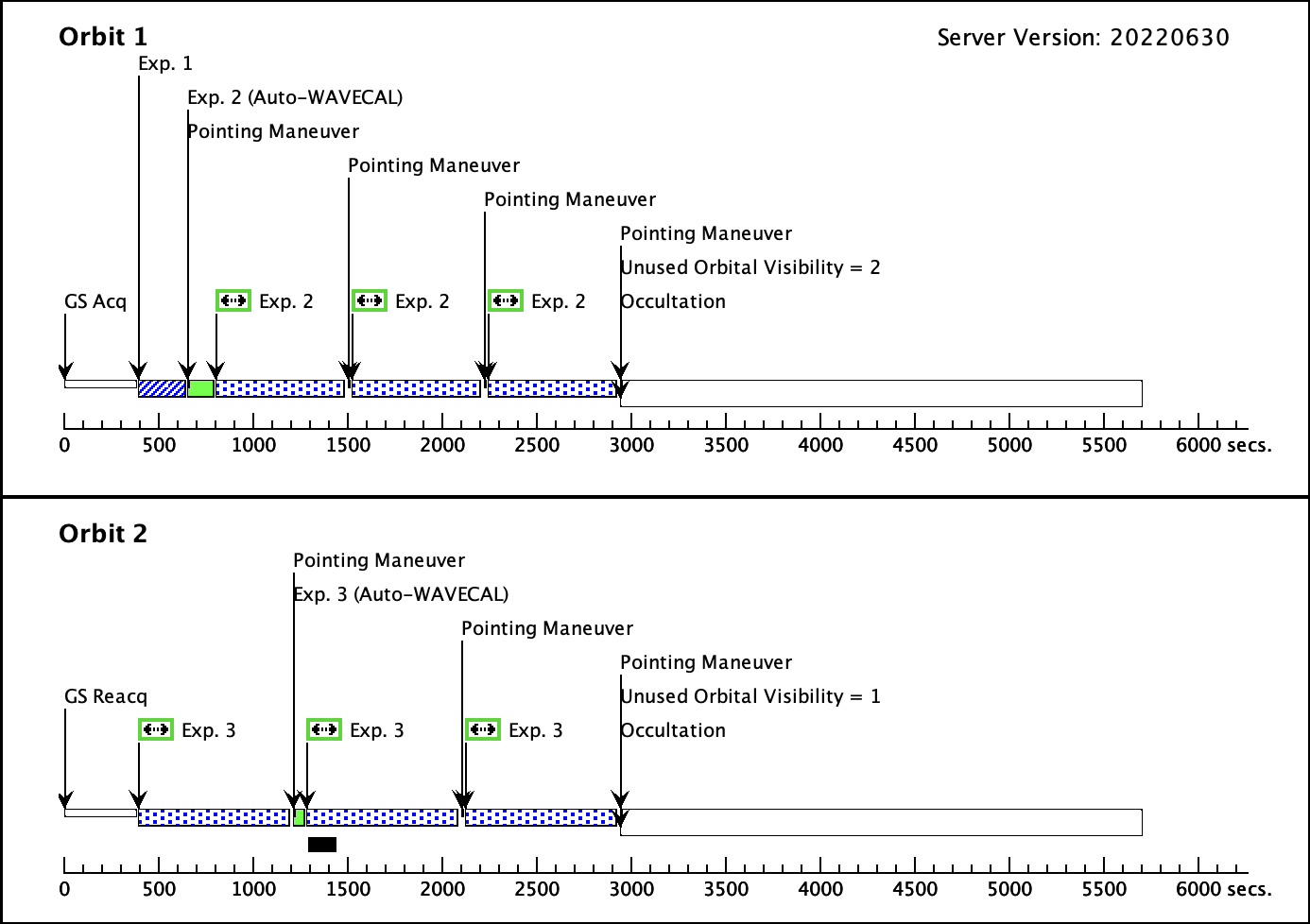}
    \caption{Image of the orbit planner tool in APT showing the planned progression of events in the first two orbits of the visit discussed in Section \ref{sec:Introduction} which experienced a failed STIS ACQ sequence. The initial guide star acquisition is placed at the beginning of the first orbit. The STIS ACQ sequence (labeled Exp. 1) is shown in the dashed blue lines following the guide star acquisition in the first orbit. This is followed by a pointing maneuver to center the target in the science aperture (the last step of the fine-locate phase). The science exposures in are shown with as dotted blue bars (labeled Exp. 2 and 3) and are sets of dithered exposures (hence the pointing maneuvers between each exposure to move the target along the detector). Note that the STIS ACQ is only present in the first orbit. When this visit was executed the initial guide star acquisition failed causing the STIS ACQ sequence and remaining exposures in the first orbit to be blank. The guide star re-acquisition in the second orbit was successful and the following science exposures were taken, but the pointing could not be guaranteed due to the missed STIS ACQ sequence.}
    \label{fig:gs_STIS_acq}
\end{figure}

We are interested in visits that have multiple orbits. At the beginning of each subsequent orbit the guide stars are reacquired and the telescope points back at the position it was at before occultation. This allows for the corrections the STIS ACQ sequence made to the pointing at the beginning of the first orbit to be carried through to any subsequent orbits. This sequence of HST guide star acquisition $\rightarrow$ STIS ACQ sequence $\rightarrow$ science exposures is shown in Figure \ref{fig:gs_STIS_acq} in the Astronomer's Proposal Tool (APT) orbit planner for the recent visit affected by a missed STIS ACQ mentioned in Section \ref{sec:Introduction}. Note that the STIS ACQ sequence is generally only performed on the first orbit following the initial guide star acquisition. The subsequent guide star re-acquisitions on later orbits maintain the pointing accuracy achieved from the STIS ACQ sequence.

\vspace{-0.3cm}
\ssection{Cataloguing STIS ACQ Success}\label{sec:catalog}

\begin{figure}[!h]
  \centering
  \includegraphics[width=\textwidth]{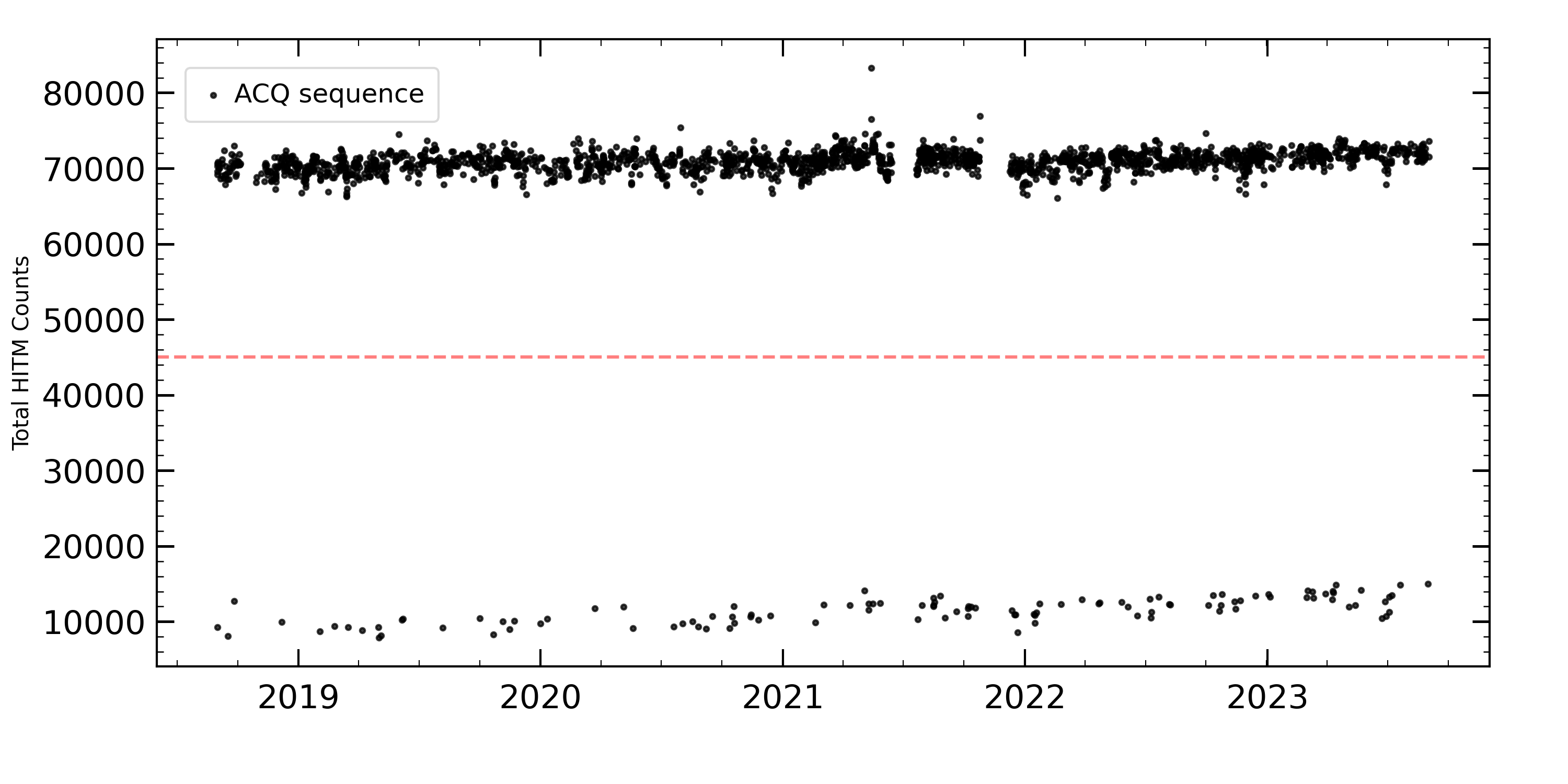}
    \caption{Total HITM counts for each STIS ACQ sequence from September 1, 2018 through September 1, 2023. The dashed line indicates the cutoff below which images were assumed to be of an unilluminated frame.}
    \label{fig:hitm_counts}
\end{figure}

In order to analyze how often visits are affected by missing STIS ACQ sequences, those failed STIS ACQ files have to be identified. To do so, we first find visits where the HITM exposure is not illuminated and the ACQ sequence images have no signal. As described in Section \ref{sec:acq}, the HITM lamp exposure is the last exposure taken in the ACQ sequence and will only occur if the take data flag (TDF) is up, which is dependent upon a successful HST guide star acquisition. As shown in Figure \ref{fig:hitm_counts}, plotting the total counts in the HITM exposure for each ACQ sequence shows two categories of images: those with an illuminated lamp and those without. We then take a cut of observations whose total HITM counts are below 45000, which constitute the subset of unilluminated HITM exposures. The time period of five years (September 1, 2018 through September 1, 2023) was chosen to get a large enough sample to perform an effective analysis while keeping it current.

The number of ACQ images with HITM counts $<$ 45000 yields 113 visits. It is possible the HST guide stars can be acquired and the ACQ images taken but the lamp illumination is delayed. In this case the ACQ sequence would be impacted but it would not be a result of guide star issues. It is therefore insufficient to rely entirely on unilluminated HITM exposures to identify missed STIS ACQ sequences as a result of HST guide star issues. As such we confirmed with each of the programmatically selected visits that there were HST guide star issues present by finding the relevant HST Exception Report. For the four visits which did not have an associated Anomaly Report (15660-03, 15463-Z4, 15517-04, and 15335-10) we verify the HST guide star issues using the STIS TDF monitor available at \href{https://www.stsci.edu/~STIS/monitors/tdf/}{https://www.stsci.edu/$\sim$STIS/monitors/tdf/}. An example of one of the selected STIS ACQ sequences which failed due to HST guide star issues to compare with the successful STIS ACQ sequence shown in Figure \ref{fig:good_acq} is given below in Figure \ref{fig:bad_acq}.

\begin{figure}[!h]
  \centering
  \includegraphics[width=\textwidth]{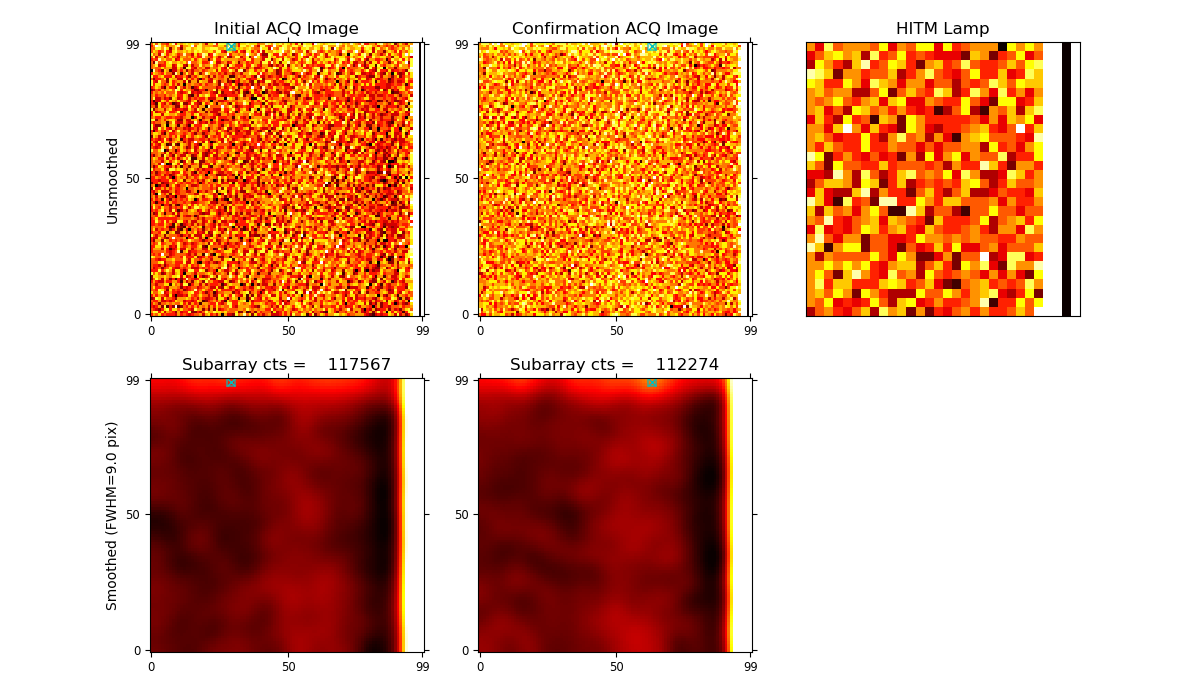}
    \caption{Images taken during the unsuccessful STIS ACQ sequence which occurred at the beginning of the visit discussed in Section \ref{sec:Introduction}. In this scenario the initial HST guide star acquisition failed at the beginning of the orbit resulting in the TDF remaining down for the STIS ACQ sequence and science exposures during that orbit.}
    \label{fig:bad_acq}
\end{figure}

\vspace{-0.3cm}
\ssection{Statistics of Guide Star impacted STIS ACQ Sequences}\label{sec:stats}

With this sample of missed STIS ACQ sequences due to guide star issues, we analyze how often they occur and if the guide stars are reacquired later on in the visit. In order to determine how often guide star issues cause STIS ACQ sequences to fail, we take the fraction of STIS ACQ sequences identified in our catalog compared to all STIS ACQ sequences taken in six-month bins. The average number of total STIS ACQ sequences for each six month bin is 214, ranging from a low of 170 in September 2019 through February 2020 to a maximum of 270 in March through August 2021. The average number of total failed STIS ACQ sequences for each six month bin is 12, ranging from a low of 6 in September 2018 through February 2019 to a maximum of 20 in March through August 2023. This is plotted below in Figure \ref{fig:failure_rate}, which shows that the rate varies between  $\sim$ 3\% to 9\% of STIS ACQ sequences. The median value is  $\sim$ 5\%, with the median number of missed to total STIS ACQ sequences being 11 to 220 per six month bin. Various issues related to pointing control system have recently been causing an increase in the amount of guide star issues and is reflected in the current high amount of missed ACQ sequences. Various mitigations are being implemented to rectify the issues. Since the pointing performance of HST fluctuates over time, the current high rate of missed STIS ACQ sequences is not necessarily indicative of the future, as seen with the similarly high rate in 2021 which subsided in 2022.  

\begin{figure}[!h]
  \centering
  \includegraphics[width=\textwidth]{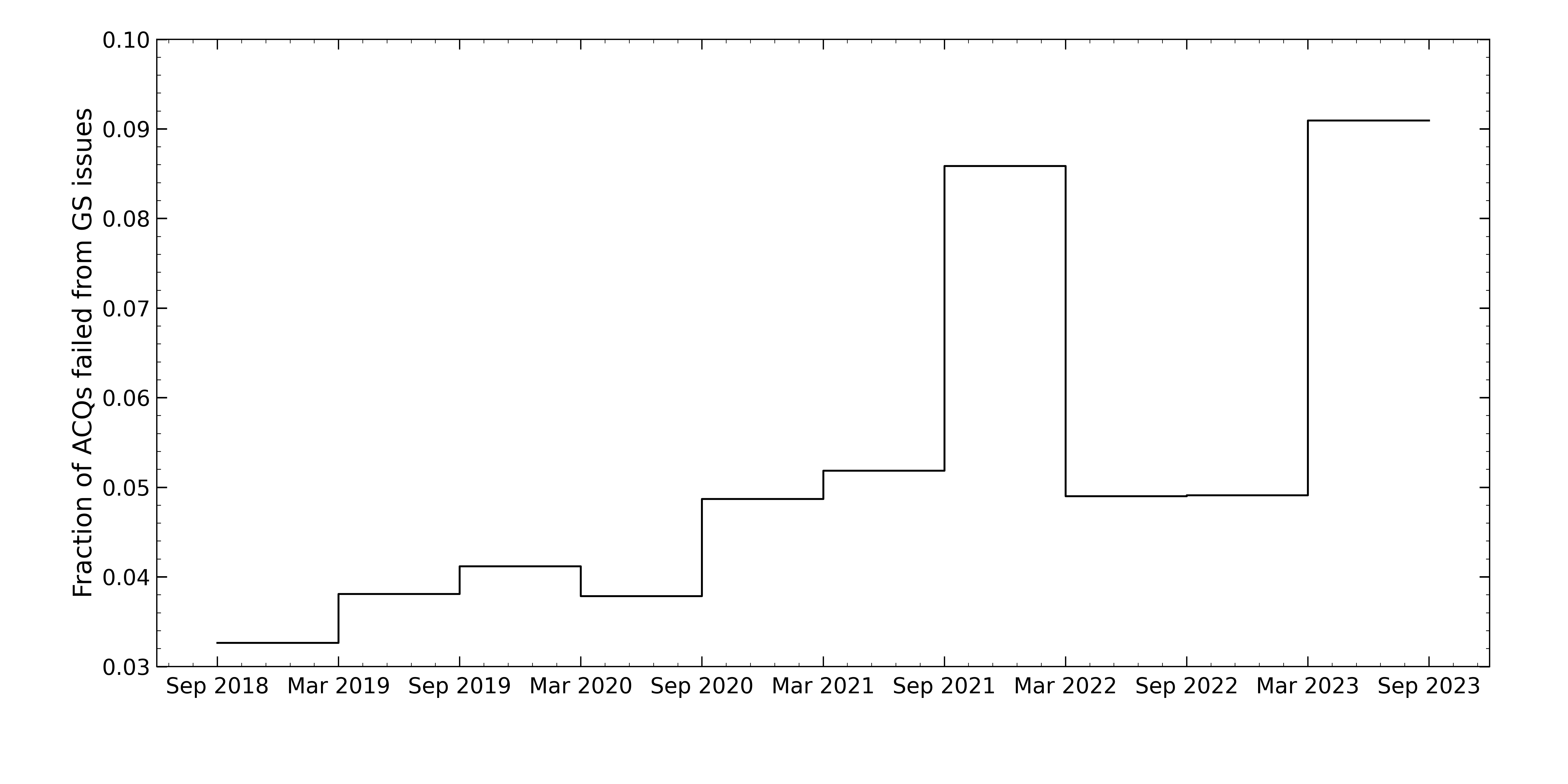}
    \caption{Failure rate of STIS ACQ sequences due to guide star issues in six month bins from September 2018 to September 2023. The failure rate has increased recently due to issues with Gyro-3, the previous increase in 2021 was due to issues with FGS-2.}
    \label{fig:failure_rate}
\end{figure}

\begin{figure}[!h]
  \centering
  \includegraphics[width=\textwidth]{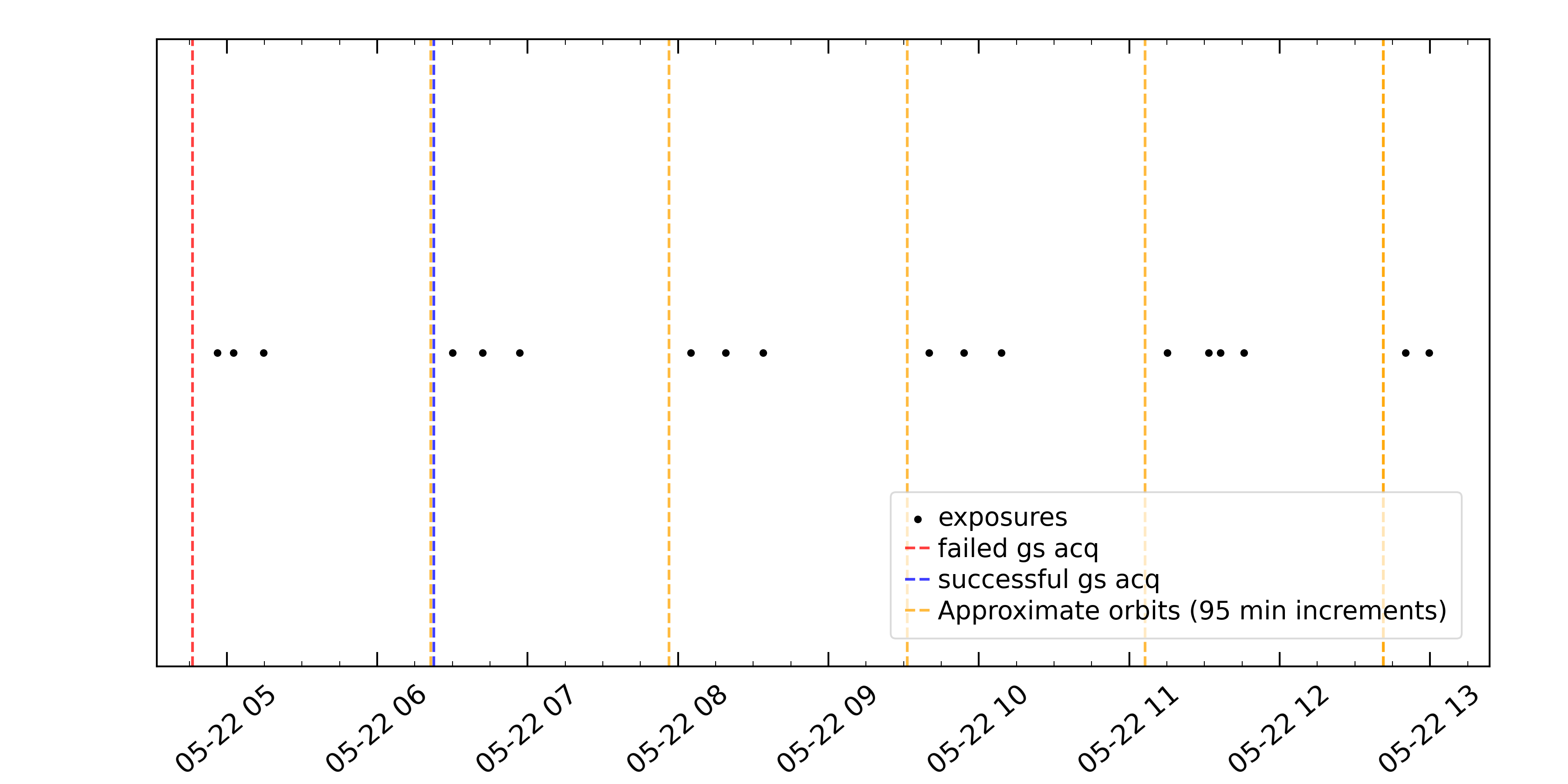}
    \caption{The start times marked as a black dot for each exposure taken during the visit discussed in Section \ref{sec:Introduction} which prompted this investigation. The times of the failed guide star acquisition, successful guide star re-acquisition, and estimate of the beginning of each orbit are marked as vertical red, blue, and orange dotted lines respectively.}
    \label{fig:17205_orbits}
\end{figure}

In order to understand when guide stars are successfully reacquired we compile the number of orbits in each selected visit, if they were reacquired, and if so on which orbit. We plot the start times of the exposures in the visit over time to get the number of orbits and overplot the times of guide star failures and re-acquisitions from the HST Exception Reports mentioned in Section \ref{sec:catalog}. An example of the methodology is shown in Figure \ref{fig:17205_orbits}, where the red vertical line is the time of the failed guide star acquisition given in the report. The blue vertical line is the timing of the successful guide star re-acquisition, also given in the report. Overplotted in orange are intervals of 95 minutes (roughly HST's orbital period) after the initial failed guide star acquisition to show where the orbits begin. The visit in Figure \ref{fig:17205_orbits} had six orbits. The guide star acquisition failed on the first orbit and the STIS ACQ sequence subsequently failed along with the two scheduled science exposures. The guide stars were then successfully reacquired in the second orbit and the remaining science exposures were taken, albeit with potentially inaccurate pointing. Each visit in the catalog, the number of orbits in the visit, and if applicable, the orbit number the guide stars were reacquired on is available in Table \ref{tab:appendix_table} in the Appendix (Section \ref{sec:appendix}).

\begin{figure}[!h]
  \centering
  \includegraphics[width=\textwidth]{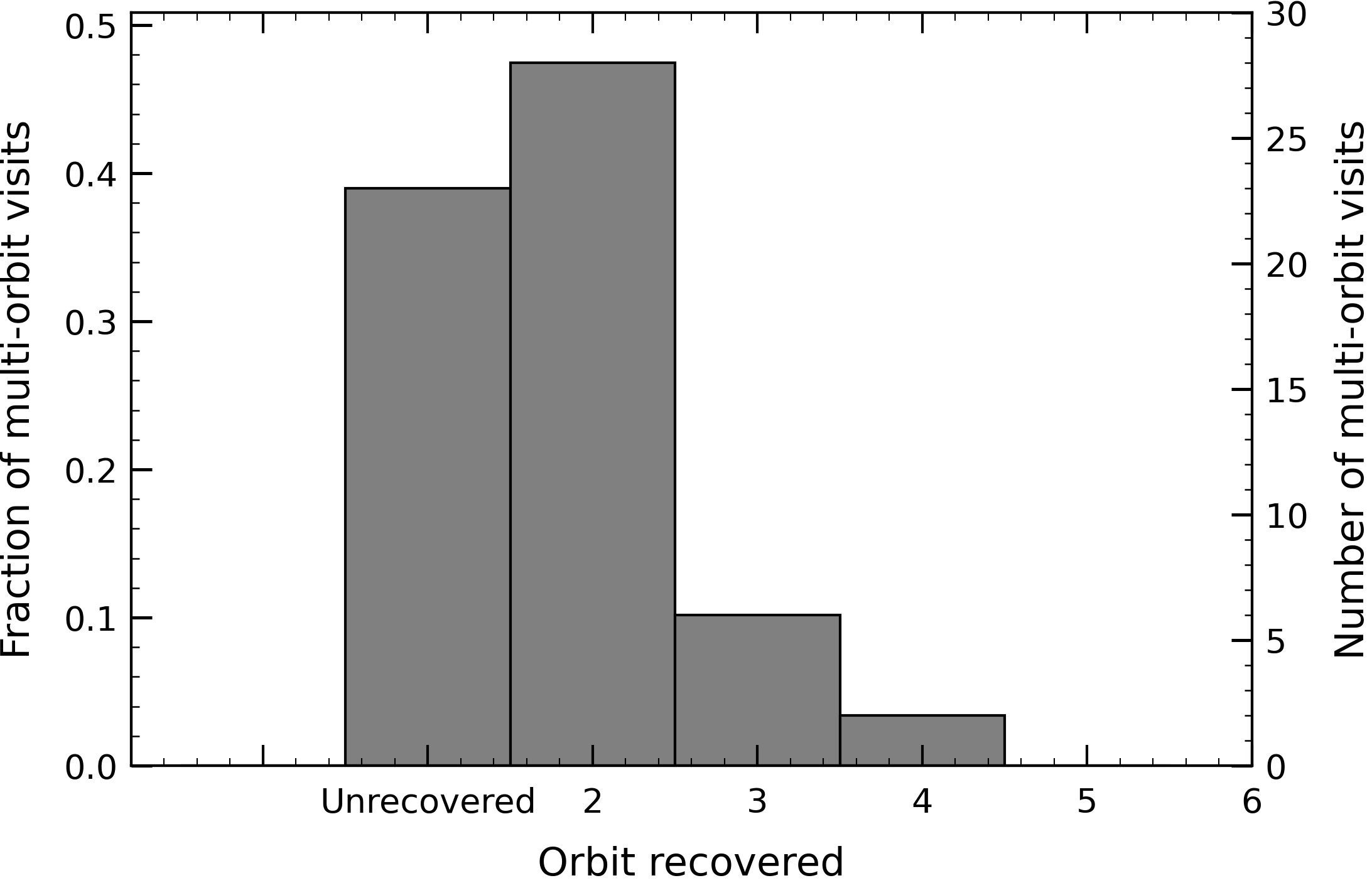}
    \caption{Histogram showing the fraction of multi-orbit visits that regain guide stars at each orbit number. There are seven visits where the guide stars were recovered in the middle of an orbit because of a delay in acquiring them or a loss of lock rather than a failure. For these, if any subsequent guide star re-acquisition were successful we count the orbit recovered to be the one on which this re-acquisition occurred since the STIS ACQ sequence was already missed for prior orbits. Placing a STIS ACQ sequence at the beginning of the orbit with the successful guide star re-acquisition would therefore improve the pointing from then on in the visit. In the three cases where the guide stars were recovered for an orbit but dropped at a later one, we count it as recovering at the earlier orbit. The reason being that if an additional STIS ACQ sequence was present it would have improved the pointing for at least the portion of the visit that had guide stars before they were lost again.}
    \label{fig:recover_rate}
\end{figure}

From this we can analyze the 59 visits that were multi-orbit in order to find if and when guide stars were reacquired. The number of orbits spanned ranges from two to six. We generate a simple histogram of the orbit number the guide stars are recovered at for these multi-orbit visits in Figure \ref{fig:recover_rate}.

Based on this histogram we note that 23 out of 59 ($\sim$ 39\%) never acquired guide stars, and for the sample of 36 visits that did recover guide stars, 28 of them ($\sim$ 78\%) were reacquired on or by the second orbit. Only 8 ($\sim$ 22\%) were recovered at any point after. 

\vspace{-0.3cm}
\ssection{Visit Planning with Safety ACQ Sequences}\label{sec:planning}

The standard procedure for repeating a visit due to observatory level failures is by filing a HOPR \footnote{\url{https://www.stsci.edu/hst/observing/post-observation/reporting-problems}}. For the majority of science cases, the timescale of the HOPR process is sufficient to reclaim lost data. However, for transient and/or rare events that are observable for a limited duration, this process may be too slow. Repeated observations of the same target requiring a $<21$ day turnaround are disruptive to the normal scheduling process and can only be accommodated on a best effort basis.

Additionally, target visibilities and other observatory constraints can limit the time frame that a failed visit could be re-scheduled in. These observations that are difficult to schedule repeat visits in the event of a failure could benefit from safety ACQ sequences. 

There are a variety of these types of science cases where safety ACQ sequences could apply. In particular, STIS Target of Opportunity (ToO) programs which observe spectroscopy of singly-occurring transient phenomena (e.g. supernovae, kilonovae, tidal disruption events, etc.) that are unique. Additionally, long duration observations ($>$5-orbit visits) with limited scheduling windows, or rare solar system events (like satellite transits or eclipses) could benefit from this technique. If a particular observation is unable to be repeated the PI may want to take advantage of the opportunity by improving the chances of success with a safety ACQ sequence at the expense of science exposure time. For example, for the supernova observation discussed in Section \ref{sec:Introduction} and throughout this work, the target was nearby and bright. In such a case using some of the science exposure time for additional safety ACQs would likely not impact the science goals.

The primary trade-off is a guaranteed loss of science exposure time to accommodate the safety ACQ sequence versus the possibility that the observation will not be able to be repeated in the case of a failed HST guide star acquisition. With this in mind, PIs should consider a number of factors when deciding whether to include a safety ACQ sequence:

\begin{itemize}
  \item The repeatability of their observations based on timing constraints.
  \item How their specific instrumental setup is affected by a failure or safety ACQ sequence. 
  \item Whether using a wider slit width can mitigate risk of scientific data loss.
  \item The overall likelihood of a guide star failure.
\end{itemize}\

The repeatability of an observation will depend on three primary factors: the timescale of the event, the rarity of the event, and the schedulability of the observations. If the timescale of the event is fast ($<21$ days) then a repeat observation is likely to disrupt the schedule. For ToO programs, PIs should consider whether a future observation of a \textit{new} event (awarded via a HOPR request) could achieve the same science. If the event is sufficiently rare (e.g., one per year) this is unlikely to be the case and a safety ACQ sequence may be desirable. This may also be preferred if at the time of Phase II submission, the target has limited schedulability. This possibility is more likely for longer-duration observations. 

The choice of instrumental setup could impact both the quality of observations in the event of a failure and the overhead if a safety ACQ sequence is performed. Primarily, consideration should be given to the specific aperture being used. The observatory pointing accuracy based on guide star positions is $\sim$ 0.35 arcseconds. This means that observations that require smaller slits or users who require extremely accurate target centering for precise wavelength calibrations are more likely to be impacted negatively by missed STIS ACQ sequences and a safety ACQ sequence may be desirable to ensure centering in the slit. Safety ACQ sequences will also alleviate any issues due to drifting even if the initial STIS ACQ sequence was successful. However, programs using slits with width $\leq 0.1$ arcsecond also require ACQ-PEAK exposures (see STIS Instrument Handbook section 8.3) and repeating these exposures in a safety ACQ sequence would significantly increase overhead. If a PI does decide to implement safety ACQs, they should work closely with their Program Coordinator to determine the best way to implement them in the APT. This could involve placing a STIS ACQ after a guide star re-acquisition, or splitting individual exposures into separate visits with an ACQ sequence at the beginning of each. Users should reach out to the helpdesk if they would like these techniques incorporated in their Phase I.  
Finally, the fraction of failed STIS ACQs has increased from 2018 to 2023. Over the last 6 months specifically it has increased to $\sim 9\%$ due to various issues related to the pointing control system. Users should discuss the most recent guiding behavior of the instrument with their program coordinators prior to planning of their observations. We note that this analysis does not generate the statistics of failures per orbit for all visits, only those where the ACQ sequence is lost due to guide star issues in the first orbit. 

\vspace{-0.3cm}
\ssection{Conclusion}\label{sec:conclusion}
Based on this work, we advise STIS observers of visits which are difficult to repeat given observatory or astrophysical constraints to consider how a failed ACQ sequence in the first orbit (caused by problems with guide star acquisition) may affect their science. By selecting the 113 STIS ACQ sequences that have experienced this issue over the past 5 years, we find that the rate at which they occur has fluctuated from $\sim$ 3\% to 9\% of all STIS ACQ sequences taken during a 6 month period. In analyzing the subset of 59 visits that spanned multiple orbits, we find that $\sim$ 39\% do not acquire guide stars at all, while $\sim$ 78\% of those that do will have them for observations taken in the second orbit. PIs should consider the possibility of their observations being impacted by guide star failures on a case by case basis. In scenarios where repeat observations are unlikely in the timescale of a HOPR it may be advantageous to include a safety ACQ sequence at the expense of time on the science target. 


\vspace{-0.3cm}
\ssectionstar{Acknowledgements}
\vspace{-0.3cm}
We thank Jenna Ryon and Sierra Gomez for their helpful discussions and suggestions regarding this project. Special thanks to Joleen Carlberg, Sean Lockwood, Serge Dieterich, Steve Arsalanian, Mike Wenz, and the Hubble Mission Office for their expertise and advice regarding analyzing STIS ACQ sequences, observation planning, and FGS behavior. 

\vspace{-0.3cm}
\ssectionstar{References}\label{sec:References}
\vspace{-0.3cm}

\noindent
Medallon, S. and Welty, D. et al. 2023 "STIS Instrument Handbook," Version 22.0, (Baltimore: STScI)
\\
Sohn, S. T., et al., 2019, “STIS Data Handbook”, Version 7.0, (Baltimore: STScI).
\\

\newpage
\vspace{-0.3cm}
\ssectionstar{Appendix: Missed ACQ Sequence Visits and Orbit Statistics}\label{sec:appendix}
\vspace{-0.3cm}

\begin{longtable}[l]{|l|c|c|c|p{3.6cm}|}
\caption{Visit information and orbit statistics for the 113 cataloged missed ACQ sequences.} \label{tab:appendix_table} \\

\hline \multicolumn{1}{|c|}{\textbf{PID - visit}} & \multicolumn{1}{c|}{\textbf{Obs date}} & 
\multicolumn{1}{c|}{\textbf{N Orbits}} & 
\multicolumn{1}{c|}{\textbf{Orbit of GS ReACQ}} & \multicolumn{1}{|c|}{\textbf{Notes}} \\ \hline 
\endfirsthead

\multicolumn{5}{c}%
{{\bfseries \tablename\ \thetable{} -- continued from previous page}} \\
\hline \multicolumn{1}{|c|}{\textbf{PID - visit}} & \multicolumn{1}{c|}{\textbf{Obs date}} & 
\multicolumn{1}{c|}{\textbf{N Orbits}} & \multicolumn{1}{c|}{\textbf{Orbit of GS ReACQ}} & \multicolumn{1}{c|}{\textbf{Notes}} \\ \hline 
\endhead

\hline \multicolumn{5}{|r|}{{Continued on next page}} \\ \hline
\endfoot

\hline \hline
\endlastfoot

15218-61  &  2018-09-01  &  1  &  N/A  &    
 \\
\hline
15327-11  &  2018-09-16  &  2  &  NONE  &    
 \\
\hline
15218-44  &  2018-09-26  &  1  &  N/A  &    
 \\
\hline
15485-02  &  2018-12-07  &  2  &  NONE  &    
 \\
\hline
15485-52  &  2019-02-02  &  2  &  NONE  &   
 \\
\hline
15463-Z4  &  2019-02-24  &  1  &  N/A  &    
 \\
\hline
15335-10  &  2019-03-17  &  1  &  N/A  &    
 \\
\hline
15093-08  &  2019-04-07  &  2  &  NONE  &    
 \\
\hline
15651-07  &  2019-05-02  &  4  &  NONE  &    
 \\
\hline
15651-08  &  2019-05-03  &  4  &  NONE  &    
 \\
\hline
15651-10  &  2019-05-06  &  1  &  N/A  &  CVZ observation. 
 \\
\hline
15517-01  &  2019-06-06  &  2  &  NONE  &    
 \\
\hline
15517-04  &  2019-06-08  &  2  &  NONE  &    
 \\
\hline
15071-31  &  2019-08-07  &  3  &  3  &    
 \\
\hline
15659-04  &  2019-10-01  &  1  &  N/A  &    
 \\
\hline
15659-12  &  2019-10-22  &  2  &  NONE  &    
 \\
\hline
15657-05  &  2019-11-05  &  2  &  NONE  &  Guide stars delayed, TDF down for acq, up for science exposure in first orbit, guide star re-acquisition failed on 2nd orbit. 
 \\
\hline
16026-02  &  2019-11-16  &  1  &  N/A  &    
 \\
\hline
15659-10  &  2019-11-23  &  1  &  N/A  &    
 \\
\hline
15815-07  &  2020-01-01  &  2  &  NONE  &    
 \\
\hline
15660-03  &  2020-01-11  &  3  &  2  &  Guide stars lost lock before and/or during acq, regained before science exposure but only up completely for 2nd orbit onward.
 \\
\hline
15653-33  &  2020-03-23  &  1  &  N/A  &    
 \\
\hline
15629-38  &  2020-05-05  &  1  &  N/A  &    
 \\
\hline
15653-54  &  2020-05-19  &  1  &  N/A  &    
 \\
\hline
15747-L3  &  2020-07-19  &  1  &  N/A  &    
 \\
\hline
15971-04  &  2020-08-02  &  1  &  N/A  &    
 \\
\hline
15419-25  &  2020-08-17  &  2  &  2  &  Guide stars lost lock before and/or during acq, regained before science exposures but only up completely for 2nd orbit.
 \\
\hline
15925-02  &  2020-08-26  &  2  &  NONE  &    
 \\
\hline
16012-82  &  2020-09-06  &  1  &  N/A  &    
 \\
\hline
15925-13  &  2020-09-16  &  1  &  N/A  &    
 \\
\hline
16230-31  &  2020-10-12  &  1  &  N/A  &    
 \\
\hline
16230-89  &  2020-10-16  &  1  &  N/A  &    
 \\
\hline
16091-2S  &  2020-10-18  &  3  &  NONE  &  Guide stars lost lock before and/or during acq, regained before science exposures in first orbit but subsequent guide star reacqs failed.
 \\
\hline
15925-23  &  2020-10-19  &  2  &  2  &    
 \\
\hline
15304-C3  &  2020-11-13  &  4  &  2  &  Guide stars lost lock before and/or during acq, regained before science exposures but only up completely for 2nd orbit onward. 
 \\
\hline
16270-02  &  2020-11-14  &  5  &  NONE  &    
 \\
\hline
16230-2J  &  2020-11-24  &  1  &  N/A  &    
 \\
\hline
16230-2V  &  2020-12-13  &  1  &  N/A  &    
 \\
\hline
16285-07  &  2021-02-19  &  2  &  2  &    
 \\
\hline
16447-15  &  2021-03-03  &  1  &  N/A  &    
 \\
\hline
16230-2H  &  2021-04-11  &  1  &  N/A  &  Guide stars lost lock before and/or during acq, regained before last science exposure. 
 \\
\hline
16230-3D  &  2021-05-04  &  2  &  NONE  &    
 \\
\hline
16225-44  &  2021-05-09  &  1  &  N/A  &    
 \\
\hline
16477-4S  &  2021-05-10  &  1  &  N/A  &    
 \\
\hline
16224-07  &  2021-05-16  &  2  &  NONE  &    
 \\
\hline
16230-93  &  2021-05-27  &  1  &  N/A  &    
 \\
\hline
15653-56  &  2021-07-23  &  1  &  N/A  &    
 \\
\hline
16263-03  &  2021-07-29  &  3  &  3  &    
 \\
\hline
16205-05  &  2021-08-15  &  1  &  N/A  &    
 \\
\hline
16205-06  &  2021-08-15  &  1  &  N/A  &    
 \\
\hline
16205-07  &  2021-08-15  &  1  &  N/A  &    
 \\
\hline
16249-AB  &  2021-08-17  &  2  &  NONE  &   
 \\
\hline
16486-01  &  2021-08-26  &  2  &  NONE  &    
 \\
\hline
16225-02  &  2021-09-02  &  1  &  N/A  &    
 \\
\hline
16368-2S  &  2021-09-19  &  3  &  2  &    
 \\
\hline
16249-9B  &  2021-10-06  &  2  &  2  &    
 \\
\hline
16249-EB  &  2021-10-06  &  2  &  NONE  &    
 \\
\hline
16166-08  &  2021-10-07  &  4  &  2  &    
 \\
\hline
16238-08  &  2021-10-11  &  2  &  2  &    
 \\
\hline
16770-01  &  2021-10-18  &  1  &  N/A  &    
 \\
\hline
16375-3S  &  2021-12-11  &  1  &  N/A  &    
 \\
\hline
16224-02  &  2021-12-16  &  4  &  3  &    
 \\
\hline
16224-09  &  2021-12-17  &  3  &  3  &    
 \\
\hline
16375-2S  &  2021-12-20  &  1  &  N/A  &    
 \\
\hline
16724-02  &  2022-01-14  &  5  &  4  &    
 \\
\hline
16708-02  &  2022-01-15  &  1  &  N/A  &    
 \\
\hline
15815-13  &  2022-01-15  &  1  &  N/A  &    
 \\
\hline
16593-AS  &  2022-01-18  &  1  &  N/A  &    
 \\
\hline
16365-5S  &  2022-01-23  &  4  &  2  &    
 \\
\hline
16218-03  &  2022-02-24  &  3  &  2  &    
 \\
\hline
16731-05  &  2022-03-27  &  1  &  N/A  &    
 \\
\hline
16772-02  &  2022-04-21  &  3  &  2  &    
 \\
\hline
16772-10  &  2022-04-23  &  2  &  2  &    
 \\
\hline
16857-1S  &  2022-05-26  &  1  &  N/A  &    
 \\
\hline
16857-2T  &  2022-06-05  &  3  &  NONE  &    
 \\
\hline
16692-10  &  2022-06-18  &  1  &  N/A  &    
 \\
\hline
16559-E3  &  2022-07-08  &  2  &  NONE  &    
 \\
\hline
16857-AS  &  2022-07-09  &  1  &  N/A  &    
 \\
\hline
16701-18  &  2022-07-10  &  4  &  2  &    
 \\
\hline
16205-57  &  2022-07-21  &  1  &  N/A  &    
 \\
\hline
16723-02  &  2022-08-05  &  5  &  3  &    
 \\
\hline
16805-1S  &  2022-08-08  &  1  &  N/A  &    
 \\
\hline
16666-21  &  2022-10-04  &  1  &  N/A  &    
 \\
\hline
17089-02  &  2022-10-11  &  1  &  N/A  &    
 \\
\hline
16701-20  &  2022-10-20  &  3  &  2  &    
 \\
\hline
16689-05  &  2022-10-23  &  6  &  2  &    
 \\
\hline
16701-21  &  2022-10-25  &  4  &  4  &    
 \\
\hline
16705-04  &  2022-11-12  &  1  &  N/A  &    
 \\
\hline
16705-05  &  2022-11-14  &  1  &  N/A  &    
 \\
\hline
16957-L1  &  2022-11-21  &  1  &  N/A  &    
 \\
\hline
17142-01  &  2022-12-14  &  2  &  2  &    
 \\
\hline
16719-01  &  2023-01-02  &  6  &  2  &  Reacq on orbit 5 delayed despite previous successful reacqs. 
 \\
\hline
16957-R1  &  2023-01-05  &  1  &  N/A  &    
 \\
\hline
16966-03  &  2023-03-01  &  1  &  N/A  &    
 \\
\hline
16655-01  &  2023-03-03  &  4  &  2  &  Reacq on orbit 4 failed despite previous successful reacqs. 
 \\
\hline
16785-01  &  2023-03-10  &  1  &  N/A  &    
 \\
\hline
16807-1S  &  2023-03-11  &  3  &  2  &    
 \\
\hline
16772-15  &  2023-03-29  &  3  &  2  &  Reacq on orbit 3 failed despite previous successful reacq. 
 \\
\hline
16966-01  &  2023-04-08  &  2  &  NONE  &    
 \\
\hline
17176-04  &  2023-04-09  &  2  &  2  &    
 \\
\hline
17169-03  &  2023-04-10  &  1  &  N/A  &    
 \\
\hline
17287-08  &  2023-04-14  &  1  &  N/A  &    
 \\
\hline
16772-17  &  2023-05-04  &  2  &  2  &    
 \\
\hline
17001-16  &  2023-05-13  &  3  &  2  &    
 \\
\hline
17205-01  &  2023-05-22  &  6  &  2  &    
 \\
\hline
16659-32  &  2023-06-23  &  2  &  2  &    
 \\
\hline
17287-15  &  2023-06-27  &  1  &  N/A  &    
 \\
\hline
17095-11  &  2023-06-28  &  1  &  N/A  &    
 \\
\hline
17166-16  &  2023-07-03  &  3  &  2  &    
 \\
\hline
17105-15  &  2023-07-04  &  4  &  3  &  2nd obrit onward taken using COS. Guide stars lost lock before and/or during
COS ACQ sequence on second orbit, regained before science exposures but only up completely for 3rd orbit onward. 
 \\
\hline
17313-04  &  2023-07-07  &  1  &  N/A  &    
 \\
\hline
16747-06  &  2023-07-20  &  3  &  NONE  &  
 \\
\hline
16700-01  &  2023-08-31  &  5  &  2  &    
 \\
\hline

\end{longtable}

\vspace{-0.3cm}
\ssectionstar{Change History for STIS ISR 2024-01}\label{sec:History}
\vspace{-0.3cm}
Version 1: 30 January 2024- Original Document 

\end{document}